# Idaho Blacks: Quiet Economic Triumph of Enduring Champions

**Rama K. Malladi**, PhD, CFA, CAIA, FRM, PMP*

Associate Professor of Finance

Department of Accounting, Finance, and Economics

College of Business and Public Policy (CBAPP)

California State University, Dominguez Hills (**CSUDH**)

1000, E Victoria Street, I&I 4320, Carson, CA 90747, USA

Email: rmalladi@csudh.edu    Phone: +1 (310) 243-3560

**Phillip Thompson** [#]

Director, The Idaho Black History Museum (IBHM)

508 Julia Davis Drive

Boise, ID 83702

Email: phillip.thompson@ibhm.org

Phone: +1 (208) 789-2164

---

*Rama K. Malladi is an associate professor of finance at CSUDH. The Watts Rebellion, also known as the Watts Riots, was a large-scale civil disturbance in the Watts and Compton neighborhoods of Los Angeles, California, in August 1965. The rioting resulted in 34 deaths, 1,032 injuries, and 3,438 arrests. CSUDH was established in 1966 within a 10-mile driving distance from Watts and Compton in the aftermath of the Watts Rebellion.

[#]Phillip Thompson is the director of the Idaho Black History Museum. The museum is a foundational pillar of the State of Idaho and its community. This year (2023) marks the 102nd anniversary of the church that housed Idaho's first Black church and now houses the IBHM.

# Idaho Blacks: Quiet Economic Triumph of Enduring Champions

**Abstract:**

As the United States (US) is witnessing elevated racial differences pertaining to economic disparities, we have found a unique example contrary to the traditional narrative. Idaho is the only US state where Blacks earn more than Whites and all other races. In this paper, we examine how Idaho Blacks might have achieved economic success and, more importantly, what factors might have led to this achievement in reducing racial and economic disparities. Preliminary research suggests that fewer barriers to land ownership, smaller populations, well-knit communities, men's involvement in the family, and a relatively less hostile environment have played a significant role. Further research by historians can help the nation uncover the underlying factors to see if some factors are transportable to other parts of the country.

# 1. INTRODUCTION

The economic gulf between Blacks and Whites that existed a half-century ago has not disappeared. Measures of household income, household wealth, and homeownership show that the gaps are as wide or wider today as they were in the 1960s and 1970s (Pew Research Center, 2013). In the US, Blacks earn 76% of Whites, according to the three-year estimates from the Current Population Survey (CPS)[1] released in July 2020 (US Dept. of Labor, 2020). The earnings trend confirms a long-known economic disparity that has persisted since the nation's founding.

However, one state stands out as an exception—Idaho. Idaho is the only US state where Blacks earn more than Whites and all other races. Asian-Pacific Islanders hold the country's top spot and are ranked fourth in Idaho. As shown in Table (1), Blacks in Idaho earn 106% of Whites and are the highest earning group. It is also interesting to note that Blacks have the lowest income disparity throughout the Pacific Northwestern states—Hawaii (ranked #2), Wyoming (#3), Montana (#4), and Oregon (#5). Though one can argue about the validity of survey results of a micro population, it is difficult to argue that Blacks in the Pacific Northwestern states have achieved near economic parity with Whites.

This paper examines how Idaho Blacks, particularly Blacks in Pacific Northwestern states, might have achieved this notable economic success. What factors may have led to this achievement in reducing racial and economic disparities? In this paper, for ease of use, we use the term Blacks though the quoted literature uses many other names such as Afro-American, African-American, Black, etc.

The flow of this paper is as follows. Section 2 details the data used in this study. Section 3 provides a historical context, and the final section contains conclusions and directions for future research.

---

[1] The CPS is sponsored jointly by the US Census Bureau and the US Bureau of Labor Statistics (BLS). It is the primary source of labor force statistics for the US population. The CPS is a monthly household survey of about 60,000 households used to gather labor-related data. It is the basis of the unemployment number published by the BLS. Summary CPS data is available at https://www.dol.gov/agencies/ofccp/about/data/earnings/race-and-ethnicity

| State | Rank | Average Weekly Earnings | Number of Workers | Percent Employed | Earnings Disparity |
|---|---|---|---|---|---|
| ID | 1 | $ 945.90 | 6,228 | 0.7% | $ 1.06 |
| HI | 2 | $ 1,082.71 | 9,998 | 1.5% | $ 0.96 |
| WY | 3 | $ 882.62 | 3,026 | 1.1% | $ 0.95 |
| MT | 4 | $ 823.49 | 2,582 | 0.5% | $ 0.95 |
| OR | 5 | $ 926.36 | 40,018 | 2.0% | $ 0.92 |
| WV | 6 | $ 721.99 | 23,610 | 3.2% | $ 0.85 |
| NE | 7 | $ 774.00 | 33,181 | 3.3% | $ 0.84 |
| NM | 8 | $ 836.39 | 16,841 | 1.9% | $ 0.83 |
| NH | 9 | $ 841.47 | 10,650 | 1.4% | $ 0.82 |
| TN | 10 | $ 752.30 | 485,233 | 15.5% | $ 0.81 |
| MO | 11 | $ 757.31 | 311,897 | 10.6% | $ 0.81 |
| WA | 12 | $ 916.07 | 133,463 | 3.7% | $ 0.80 |
| KY | 13 | $ 700.28 | 156,407 | 7.9% | $ 0.80 |
| IN | 14 | $ 730.75 | 288,170 | 8.8% | $ 0.79 |
| MI | 15 | $ 759.47 | 544,634 | 11.6% | $ 0.79 |
| OK | 16 | $ 748.21 | 116,705 | 6.6% | $ 0.78 |
| AK | 17 | $ 879.66 | 9,411 | 2.8% | $ 0.78 |
| MD | 18 | $ 988.30 | 850,971 | 27.5% | $ 0.78 |
| AR | 19 | $ 690.69 | 190,401 | 14.6% | $ 0.78 |
| OH | 20 | $ 717.16 | 589,908 | 10.7% | $ 0.77 |
| DE | 21 | $ 764.75 | 98,818 | 21.2% | $ 0.76 |
| GA | 22 | $ 786.14 | 1,490,304 | 30.4% | $ 0.76 |
| **NATIONAL** | | **$ 791.01** | **17,781,185** | **11.4%** | **$ 0.76** |
| MS | 23 | $ 684.03 | 431,253 | 35.4% | $ 0.75 |
| UT | 24 | $ 694.40 | 20,532 | 1.3% | $ 0.75 |
| IL | 25 | $ 818.17 | 709,543 | 11.5% | $ 0.75 |
| WI | 26 | $ 696.10 | 139,228 | 4.6% | $ 0.74 |
| IA | 27 | $ 672.68 | 50,542 | 3.0% | $ 0.74 |
| TX | 28 | $ 826.46 | 1,595,717 | 12.0% | $ 0.74 |
| NJ | 29 | $ 890.98 | 553,553 | 12.8% | $ 0.74 |
| PA | 30 | $ 731.91 | 579,153 | 9.4% | $ 0.74 |
| AL | 31 | $ 705.75 | 527,129 | 24.8% | $ 0.74 |
| NY | 32 | $ 839.10 | 1,217,299 | 13.2% | $ 0.74 |
| SC | 33 | $ 719.48 | 548,928 | 24.3% | $ 0.73 |
| NV | 34 | $ 732.42 | 121,478 | 8.3% | $ 0.73 |
| KS | 35 | $ 697.86 | 67,323 | 4.7% | $ 0.73 |
| RI | 36 | $ 770.26 | 30,425 | 5.7% | $ 0.73 |
| NC | 37 | $ 738.42 | 988,384 | 20.6% | $ 0.72 |
| VA | 38 | $ 843.18 | 774,913 | 18.3% | $ 0.72 |
| AZ | 39 | $ 761.83 | 143,294 | 4.4% | $ 0.72 |
| CA | 40 | $ 910.49 | 952,215 | 5.1% | $ 0.72 |
| FL | 41 | $ 723.17 | 1,479,322 | 15.0% | $ 0.72 |
| ND | 42 | $ 682.29 | 12,094 | 3.0% | $ 0.72 |
| MA | 43 | $ 849.90 | 274,685 | 7.5% | $ 0.72 |
| MN | 44 | $ 755.94 | 154,602 | 5.2% | $ 0.71 |
| CO | 45 | $ 812.40 | 103,329 | 3.5% | $ 0.71 |
| CT | 46 | $ 802.80 | 177,293 | 9.6% | $ 0.70 |
| LA | 47 | $ 684.75 | 575,170 | 28.7% | $ 0.69 |
| VT | 48 | $ 626.87 | 4,136 | 1.2% | $ 0.67 |
| ME | 49 | $ 562.99 | 6,011 | 0.9% | $ 0.63 |
| DC | 50 | $ 1,041.05 | 123,067 | 32.0% | $ 0.61 |
| SD | 51 | $ 525.69 | 8,114 | 1.8% | $ 0.60 |

**Table 1: Statewide Average Weekly Earnings and Rank, Number of Black Workers, and Earnings Disparity**

- Source: Earnings Disparities by Race and Ethnicity, based on the Current Population Survey (CPS) (US Dept. of Labor, 2020). Date range: January 2017 – December 2019.
- Race or Ethnicity: The race/ethnicity variable includes six categories: 1) White; 2) Black; 3) Native American/American Indian; 4) Asian-Pacific Islander; 5) Hispanic/Latino; and 6) Multiracial.
- Average Weekly Earnings: Weekly earnings include overtime, commission, and tips but exclude self-employed earnings. The average is the weighted mean.
- Number of Workers: This number includes all full-time and part-time workers.
- Percent Employed: These are the percentages of all workers who are White, Black, Native American, Asian-Pacific Islander, Hispanic/Latino, and Multiracial.
- Earnings Disparity: This is the mean weekly earnings for Black, Native American/American Indian, Asian-Pacific Islander, Hispanic/Latino, or Multiracial workers divided by the mean weekly earnings for White workers.
- Observation #1: Blacks in Idaho earn 106% of Whites and are the highest earning group.
- Observation #2: Blacks have the lowest income disparity throughout the Pacific Northwestern states; Hawaii (ranked #2), Wyoming (#3), Montana (#4), and Oregon (#5).
- Observation #3: The top-five states with the lowest Black income disparity have less than 2% Blacks in the working-age populations.

## 2. DATA

The data for this project is obtained from the following five sources:
   a) Earnings disparity data is based on the Current Population Survey (CPS) three-year dataset from January 2017 to December 2019. The CPS is a monthly household survey of about 60,000 households used to gather labor-related data. It is the basis of the unemployment number published by BLS. The CPS earnings include overtime, commission, and tips but exclude self-employed earnings. The Earnings Disparity Data contains two tables per state and two national tables.
   URL: https://www.dol.gov/agencies/ofccp/about/data/earnings/race-and-ethnicity
   b) Population data from 1870 to 1990 is available in the 'Historical Census Statistics on Population Totals by Race from 1790 to 1990' (Gibson & Jung, 2002).
   c) Population Data for 2000 from Idaho Ebony (Oliver, 2006).
   d) Population data for 2010 and 2020 is obtained from the Decennial Census Redistricting.
   URL: https://data.census.gov/table?g=0400000US16&tid=DECENNIALPL2020.P1
   e) American Community Survey (ACS) 2021 is used for household type analysis.
   URL: https://data.census.gov/table?q=B11001B&tid=ACSDT1Y2021.B11001B

## 3. ECONOMIC HISTORY OF BLACKS IN IDAHO

When Congress created Idaho Territory in 1863, the new territory sprawled across an area one-quarter larger than Texas. Today's state is smaller, but Idaho is still as large as all six New England states combined with New Jersey, Maryland, and Delaware. Consequently, to travel from Bonner's Ferry in the north to Montpelier in the southeast encompasses nearly 800 miles, only slightly less than traveling from New York City to Chicago.[2]

According to the US Census Bureau's population estimate in July 2022, approximately 1% of Idaho's population is reported as 'Black or African American alone.'[3] This contrasts the 14% national average and 2.5% in South Dakota, which has the nation's highest earnings disparity, where Blacks earn 60% of Whites. The state's population of 1.9 million ranks 38 (out of 51) in the nation. Out of this, the Black population of 20,355 ranks 46 (out of 51) in the nation out of a total Black population of 47 million (Moslimani et al., 2023).

---

[2] Idaho has a Rich and Fascinating History: https://www.idaho.gov/about-idaho/history/

[3] Comparision of ID, SD, and US populations: https://www.census.gov/quickfacts/fact/table/SD,ID,US/

Though a micro minority, Blacks have been part of Idaho's history from the inception of the current state. As shown in Table (2), in 1870, there were 60 Blacks in the Idaho territories (Gibson & Jung, 2002). By 2020, the Black population had reached 15,726 or increased by 262 times; in comparison, the state's overall population increased by 123 times and the White population by 142 times during the same period. As shown in Table (3), the Black population has grown in double-digit percentages in all decades except during the era of the great depression and world wars.

| Census year | Total population | Race | | | | |
|---|---|---|---|---|---|---|
| | | White | Black | Native Indian | Asian/Pacific | Other |
| 2020 | 1,839,106 | 1,510,360 | 15,726 | 25,621 | 30,562 | 256,837 |
| 2010 | 1,567,582 | 1,396,487 | 9,810 | 21,441 | 21,386 | 118,458 |
| 2000 | 1,293,953 | 1,177,304 | 5,456 | 10,725 | 11,889 | 88,579 |
| 1990 | 1,006,749 | 950,451 | 3,370 | 13,780 | 9,365 | 29,783 |
| 1980 | 943,935 | 901,641 | 2,716 | 10,521 | 5,948 | 23,109 |
| 1970 | 712,567 | 698,802 | 2,130 | 6,687 | 3,212 | 1,736 |
| 1960 | 667,191 | 657,383 | 1,502 | 5,231 | 2,758 | 317 |
| 1950 | 588,637 | 581,395 | 1,050 | 3,800 | 2,304 | 88 |
| 1940 | 524,873 | 519,292 | 595 | 3,537 | 1,449 | (X) |
| 1930 | 445,032 | 438,840 | 668 | 3,638 | 1,886 | (X) |
| 1920 | 431,866 | 425,668 | 920 | 3,098 | 2,180 | (X) |
| 1910 | 325,594 | 319,221 | 651 | 3,488 | 2,234 | (X) |
| 1900 | 161,772 | 154,495 | 293 | 4,226 | 2,758 | (X) |
| 1890 | 88,548 | 82,117 | 201 | 4,223 | 2,007 | (X) |
| 1890 | 84,385 | 82,018 | 201 | 159 | 2,007 | (X) |
| 1880 | 32,610 | 29,013 | 53 | 165 | 3,379 | (X) |
| 1870 | 14,999 | 10,618 | 60 | 47 | 4,274 | (X) |

**Table 2: Idaho State Population from 1870 to 2020.**
- Note: 1870 to 1990 data is available in the 'Historical Census Statistics on Population Totals by Race from 1790 to 1990' (Gibson & Jung, 2002). Data for 2000 is obtained from Idaho Ebony (Oliver, 2006). 2010 and 2020 data is obtained from the Redistricting Data below: https://data.census.gov/table?g=0400000US16&tid=DECENNIALPL2020.P1
- For comparison purposes, data under the column "Other" is not segmented further.
- Observation #1: In 1870, there were 60 Blacks in Idaho. By 2020, the Black population had reached 15,726 or increased by 262 times.
  Observation #2: Idaho's total population increased by 123 times and the White population by 142 between 1870 and 2020.

| Census year | Total population | Race | | | | |
|---|---|---|---|---|---|---|
| | | White | Black | Native Indian | Asian/Pacific | Other |
| 2020 | 17% | 8% | 60% | 19% | 43% | 117% |
| 2010 | 21% | 19% | 80% | 100% | 80% | 34% |
| 2000 | 29% | 24% | 62% | -22% | 27% | 197% |
| 1990 | 7% | 5% | 24% | 31% | 57% | 29% |
| 1980 | 32% | 29% | 28% | 57% | 85% | 1231% |
| 1970 | 7% | 6% | 42% | 28% | 16% | 448% |
| 1960 | 13% | 13% | 43% | 38% | 20% | 260% |
| 1950 | 12% | 12% | 76% | 7% | 59% | |
| 1940 | 18% | 18% | -11% | -3% | -23% | |
| 1930 | 3% | 3% | -27% | 17% | -13% | |
| 1920 | 33% | 33% | 41% | -11% | -2% | |
| 1910 | 101% | 107% | 122% | -17% | -19% | |
| 1900 | 83% | 88% | 46% | 0% | 37% | |
| 1890 | 5% | 0% | 0% | 2556% | 0% | |
| 1890 | 159% | 183% | 279% | -4% | -41% | |
| 1880 | 117% | 173% | -12% | 251% | -21% | |
| 1870 | | | | | | |

**Table 3: Idaho State Population percent changes from 1870 to 2020.**
- Note: The data in this table is based on Table (2) data.
- Percentage change = (population in the current cycle/population in the previous cycle – 1 ).
- Observation #1: Black population grew in double-digit percentages in all decades except during the era of the great depression and world wars.

A detailed historical narrative of Idaho Blacks is, unfortunately, sparse. All that exists is a collection of newspaper articles, census data, and some anecdotes. Most academic literature on Idaho Blacks centers around the works of a few authors who focused on the West in general or the Pacific Northwest (Oliver, 2006; Taylor, 1977, 1999, 2009). Many historians quote the arrival of the famous Buffalo Soldiers from the 25th infantry in 1892, sent by President Benjamin Harrison into northern Idaho to reestablish order following an outbreak of violence by striking White miners (Taylor, 2009). By the 1890s, Black soldiers in the West had assumed a new responsibility in the growing labor struggles of the region (Taylor, 1999).

In 1890, out of a total US population of 62.6 million, 35% were engaged in gainful occupations. Blacks comprised 7.5 million; however, their workforce participation rate was much greater (41%) than the total population (Gannett, 1895). Within Blacks, females comprised 32% of the workforce. Similarly, Idaho reported 106 Black workers (83 or 78% male and 23 or 22% female).

An estimated 78 were employed in domestic and personal service; 17 in agriculture, fisheries, and mining; 8 in trade and transportation; and 3 in manufacturing and mechanical industries (Gannett, 1895). In comparison, after 70 years (in 1960), females comprised 36% of the Black workforce in Idaho (Nicol & Keeler, 1964). Fast forward another 68 years, females comprised 37% of the Black workforce in Idaho in 2018, according to the 2014-2018 equal employment opportunity data (US Census Bureau, 2018).

Once many Blacks migrated to a town or city, the first community institution they established was usually a church. The church sponsored or fostered other activities such as social clubs, literary societies, and fraternal orders. The church building was often used as a meeting place for these organizations and political gatherings. Local Black ministers became the community leaders and spokesmen. The largest and most successful organization was the Prince Hall Masonic fraternal order. The origin and practice of freemasonry among colored men in the US came direct from the Mother Grand Lodge of the world, London, England, in 1784 (Grimshaw et al., 1903). The first Masonic lodge in Idaho was opened in 1899 with 48 colored members (Grimshaw et al., 1903). The Freemasons kept a detailed log of the number of enslaved people by state, the number of people of color, literacy rates, and their wealth. As of 1900, Freemasons documented 293 Idaho Blacks who owned 36 homes and 545 farms (Grimshaw et al., 1903). The farmland ownership of Idaho Blacks was the 3$^{rd}$ highest in the country after South and North Dakota (Grimshaw et al., 1903). By 1903 the various lodges which had previously been affiliated with masonic organizations in Illinois and Iowa formed the Grand Lodge of Washington. Masonic lodges in Idaho and Montana joined the Grand Lodge of Washington, which by 1920 included thirty-five lodges (Taylor, 1977).

Idaho's newspaper, 'The Gem State Rural,' was founded on 15 September 1895.[4] It was a monthly publication that focused on the lives and experiences of Idaho Blacks and the Pacific Northwest region of the United States. The newspaper aimed to address issues of discrimination and inequality that Blacks faced in the state and to provide a voice for the community. The publication ceased operations in 1911 due to financial difficulties. Many newspapers in the state have online archives and document historical events as they happened. For example, the Coeur d'Alene labor strike is narrated under the front page "Martial Law!" section of the 16th July 1892 newspaper.[5] Going through these digitized archives provides great information for academic researchers. We could identify 65 newspapers with digital archives.

---

[4] 1$^{st}$ day's photocopy of Idaho's first newspaper: https://www.newspapers.com/image/888291825/

Job opportunities for Blacks that evolved in small cities lasted through World War I but disappeared during the depression. This forced many Blacks in the towns to seek employment in Seattle or Portland or leave the region. To many Blacks, towns like Boise, Butte, or even Spokane would be temporary stops on their way to the larger cities on the coast. For others, the difficulty of finding employment in smaller towns forced a move to larger communities (Taylor, 1977). As shown in Table (3), the Black population in Idaho declined by double-digit percentages during the era of the great depression and world wars. The great depression era had a bigger negative impact on Black farmers in the West. Though the acreage under the White farmers increased by 44.7 million acres (or 26% from 1920 to 1930), it was reduced by 1.6% for Black farmers (US Census, 1933). The dollar value of farmland owned by Blacks decreased by 21% in this decade.

One area of distinction for the Idaho Black farmers was their investments in the farmland. The Homestead Act, enacted during the Civil War in 1862, provided that any adult citizen (including freed slaves), or intended citizen, who had never borne arms against the U.S. government could claim 160 acres of surveyed government land. Claimants were required to live on and improve their plots by cultivating the land. After five years on the land, the original filer was entitled to the property, free and clear, except for a small registration fee (Homestead Act, 1862). Homestead Act, which remained active for 124 years until it was repealed in 1976, resulted in 10 percent of U.S. land—or 270 million acres—being claimed and settled (History com, 2022). By 1910, Blacks in Idaho acquired and operated 1,043 acres of farmland. Black farmers in Idaho operated advanced machinery, including automobiles, and consistently ranked second in the nation. The 1930 national average value of implements and machinery per acre operated by Blacks was $109. Nebraska ranked first with an average value of $1,324, followed by Idaho, Montana, and Minnesota, with average values of $961, $959, and $920, respectively (US Census, 1933).

A study of Idaho's position compared to other western states indicates the relative prosperity of the people of this commonwealth (Idaho University School of Business Administration, 1928). As shown in Table (4), in 1921, Idaho was relatively prosperous compared to the other Western states. In Group I (lowest annual income group), those with an income of $1,000 to $2,000 constitute only 33 percent of the income taxpayers of Idaho, whereas a larger percentage is found in practically all the Western states: Arizona, 34%; California, 37%; Montana, 44%; Nevada, 41%; Oregon, 36%; Utah, 39%; Washington, 42%; Wyoming, 41%. The upper three income groups (IV to VI) constituted 13% in Idaho, the highest of all Western states. However, as shown in Table (4), by 2021, the other Western states will have outperformed in the higher income deciles. Idaho lost its prominence mainly due to the technological and industrial progress of states like Washington.

---

[5] The Coeur d'Alene, 16th July 1892: https://www.newspapers.com/image/893105129/

Political factors also played a significant role in determining the economic success of the Idaho community. In this regard, it can be said that, unlike other black communities of the time, Idaho Blacks did not encounter significant mob lynching. Between 1882 and 1946, more than two and one-half times as many Blacks as Whites were lynched in the US (3,425 Blacks vs. 1,291 Whites). Mississippi had the highest number of lynching for the South, 574, and the highest for the United States. Idaho had 20 White lynching but none for Blacks (Work & Tuskegee Institute, 1914). Therefore, it is logical to infer that the social stability in Idaho provided a great environment for farming and businesses, increasing the household income of many Idahoan Blacks.

The large-scale migration of Blacks to the Pacific Northwest during and immediately after World War II continued into the 1950s. However, the growing Black population seemed to strengthen the resolve of many white labor officials, employers, and residential associations to keep the Blacks out. Several unions in the Pacific Northwest during the 1950s discriminated against Black workers. The most serious problem facing Pacific Northwest Blacks was discrimination in housing. While discrimination in employment and public accommodations remained constant or slightly diminished, the concentration of Blacks in substandard housing in the oldest sections of the cities increased in the 1950s and early 1960s (Taylor, 1977).

Idaho Blacks did not appear to focus on the military for employment. For example, in 1970, only one Idaho Black was employed in the Army and Air National Guard (out of 4,257 from the state total). The percentage participation rate of Idaho Blacks was the lowest of all US states (US Office of the Deputy Assistant Secretary of Defense, 1971).

Idaho Blacks also did not seek political leadership as late as the 1970s. There were 2,264 elected Black officials in the United States in 1972. Most held offices in city government or law enforcement and educational-type positions. The States with the largest number of Black officials were New York, New Jersey, Ohio, Illinois, Michigan, North Carolina, Mississippi, Louisiana, and California; each had at least 100 officials. However, few states had none, and Idaho was one, along with North and South Dakota, Maine, Montana, and Utah (McKenney, 1972). In 1973, Pocatello city elected Thomas "Les" Purce to its city council - making him the first Black person to serve in public office in Idaho. Purce went on to become the Mayor of the city.[6] Idaho Blacks' participation in graduate and professional education also remained surprisingly low. As late as 1976, Idaho granted only one master's degree to a Black student out of 615 the states' total (Lehner & United States, 1981).

---

[6] Purce: https://www.nytimes.com/1976/01/03/archives/pocatello-official-becomes-idahos-first-black-mayor.html

| Deciles | 1921 Data | | | | | | | | | | | |
|---|---|---|---|---|---|---|---|---|---|---|---|---|
| | Arizona | California | Colorado | Idaho | Montana | Nevada | New Mexico | Oregon | Utah | Washington | Wyoming | Total US |
| I | 34 | 37 | 35 | 33 | 44 | 41 | 30 | 36 | 39 | 42 | 41 | 34 |
| II | 42 | 30 | 40 | 35 | 38 | 41 | 45 | 35 | 34 | 36 | 30 | 34 |
| III | 11 | 14 | 13 | 15 | 8 | 7 | 13 | 13 | 12 | 10 | 10 | 14 |
| Sum (I to III) | 87 | 81 | 88 | 83 | 90 | 89 | 88 | 84 | 85 | 88 | 81 | 82 |
| IV | 4 | 7 | 6 | 8 | 4 | 4 | 7 | 7 | 6 | 4 | 5 | 7.3 |
| V | 2 | 3 | 2 | 3 | 2 | 3 | 1.5 | 3 | 2 | 2 | 3 | 2.9 |
| VI | 0 | 2 | 1 | 2 | 1 | 0 | 0 | 2 | 0 | 1 | 2 | 1.8 |
| Sum (IV to VI) | 6 | 12 | 9 | 13 | 7 | 7 | 9 | 12 | 8 | 7 | 10 | 12 |

| Deciles | 2021 Data | | | | | | | | | | | |
|---|---|---|---|---|---|---|---|---|---|---|---|---|
| | Arizona | California | Colorado | Idaho | Montana | Nevada | New Mexico | Oregon | Utah | Washington | Wyoming | Total US |
| I | 11 | 11 | 11 | 12 | 13 | 12 | 14 | 11 | 12 | 9 | 12 | 13 |
| II | 19 | 18 | 16 | 18 | 18 | 20 | 21 | 17 | 17 | 14 | 16 | 18 |
| III | 26 | 23 | 23 | 24 | 24 | 27 | 25 | 24 | 23 | 23 | 22 | 24 |
| Sum (I to III) | 56 | 52 | 50 | 55 | 56 | 58 | 61 | 53 | 52 | 46 | 51 | 55 |
| IV | 14 | 13 | 15 | 15 | 14 | 14 | 13 | 14 | 15 | 15 | 15 | 14 |
| V | 9 | 9 | 10 | 10 | 9 | 8 | 8 | 10 | 10 | 10 | 10 | 9 |
| VI | 13 | 14 | 16 | 13 | 13 | 12 | 11 | 15 | 15 | 17 | 16 | 14 |
| Sum (IV to VI) | 35 | 36 | 40 | 38 | 37 | 34 | 32 | 39 | 40 | 42 | 41 | 36 |

**Table 4: Personal Income Tax Distributions of Western US States in 1921 and 2021.**

- Note: Unlike the 2021 data, the 1921 data had no decile for zero adjustable gross income (Idaho University School of Business Administration, 1928).
- The table includes only the first six non-zero deciles for comparison purposes.
- Observation #1: In 1921, Idaho was relatively prosperous compared to the other Western states. In Group I (lowest annual income group), those with an income of $1,000 to $2,000 constitute only 33 percent of the income taxpayers of Idaho, whereas a larger percentage is found in practically all the Western states: Arizona, 34%; California, 37%; Montana, 44%; Nevada, 41%; Oregon, 36%; Utah, 39%; Washington, 42%; Wyoming, 41%.
- Observation #2: In 1921, the upper three income groups (IV to VI) constituted 13% in Idaho, the highest of all Western states.
- Observation #3: By 2021, the other Western states have outperformed in the higher income deciles, and Idaho lost its prominence.

The family as an institution has been strong in Idaho. In 1980, the majority—56 percent—of the 6.0 million Black families were maintained by married couples nationally. This family type prevailed despite the dramatic increase between 1970 and 1980 in the proportion of Black families with a female householder and no husband present, which rose from 28 to 38 percent of total Black families over the ten years. Major factors associated with the increase since 1970 in the proportion of Black women maintaining families have been the rapid gain in divorce and separation and the rise in the number of never-married Black mothers who maintain their own families. Wisconsin had the highest proportion of its Black families maintained by a woman—48 percent-a figure eight times higher than the corresponding level for North Dakota, reporting the lowest proportion (6 percent). Idaho, with 13%, is the sixth lowest in the country (U.S. Dept. of Commerce & The Census Bureau, 1987).

The Idaho Blacks' household type is characteristically different from the Blacks in the country in one aspect—the single-family female householder proportion is significantly less in Idaho than in the country (6% in Idaho vs. 26% in the US), as shown in Table (5). In addition, the single-family male householder proportion is significantly more in Idaho than in the country (22% in Idaho vs. 6% in the US). The other family types are similar in Idaho and the country. Besides, Whites live in family households more than Blacks (4% more in the US and 15% more in Idaho). The most common household type for Whites is a married couple living together (50% in the US and 56% in Idaho). The most common household type for Blacks is living alone (36% in the US and 39% in Idaho).

| 2021 Data - Household Type | Idaho Blacks | | Idaho Whites | | USA Blacks | | USA Whites | |
|---|---|---|---|---|---|---|---|---|
| Total: | 4,178 | 100% | 591,223 | 100% | 15,274,070 | 100% | 85,344,506 | 100% |
| Family households: | 2,212 | 53% | 402,977 | 68% | 9,020,360 | 59% | 54,049,341 | 63% |
| Married-couple family | 1,033 | 25% | 333,345 | 56% | 4,118,941 | 27% | 43,067,890 | 50% |
| Other family: | 1,179 | 28% | 69,632 | 12% | 4,901,419 | 32% | 10,981,451 | 13% |
| Male householder, no spouse present | 918 | *22%* | 23,649 | 4% | 930,829 | *6%* | 3,455,880 | 4% |
| Female householder, no spouse present | 261 | *6%* | 45,983 | 8% | 3,970,590 | *26%* | 7,525,571 | 9% |
| Nonfamily households: | 1,966 | 47% | 188,246 | 32% | 6,253,710 | 41% | 31,295,165 | 37% |
| Householder living alone | 1,645 | 39% | 146,549 | 25% | 5,463,135 | 36% | 24,922,569 | 29% |
| Householder not living alone | 321 | 8% | 41,697 | 7% | 790,575 | 5% | 6,372,596 | 7% |

**Table 5:** Household Type Comparisons between Blacks and Whites in Idaho and the US in 2021.
- Note: The American Community Survey (2021) data from https://www.census.gov/programs-surveys/acs/ is summarized below.
- The table includes only the number and percentages of households in Idaho and the US for Blacks and Whites
- Observation #1: Whites live in family households more than Blacks (4% more in the US and 15% more in Idaho).
- Observation #2: Whites' most common household type is a married couple living together (50% in the US and 56% in Idaho).
- Observation #3: The most common household type for Blacks is living alone (36% in the US and 39% in Idaho).
- Observation #4: Idaho Blacks significantly differ from the Blacks in the country in one area: The single-family female householder proportion is significantly less in Idaho than in the country (6% in Idaho vs. 26% in the US)
- Observation #5: The single-family male householder proportion is significantly more in Idaho than in the country (22% in Idaho vs. 6% in the US). The other family types are similar in Idaho and the country.

Blacks have the lowest income disparity throughout the Pacific Northwestern states, where the Black population percentages are less than 2%. As shown in Figure (1), the slope is negative between the earnings disparity and the percentage of Black workers in the workforce, indicating that as the earnings disparity grows, the percentage of Blacks increases. Both the coefficient and constant are statistically significant at the 5% level. For every 10% increase in the percentage of Blacks in the workforce, the earnings disparity goes up by 2.9%. It is difficult to pinpoint what exactly caused the economic success of Idaho Blacks without further research. Based on anecdotal stories and limited historical data, we offer a conjecture that Blacks economically prospered in environments with fewer barriers to land ownership, smaller populations, well-knit communities, men's involvement in the family, and a relatively less hostile environment.

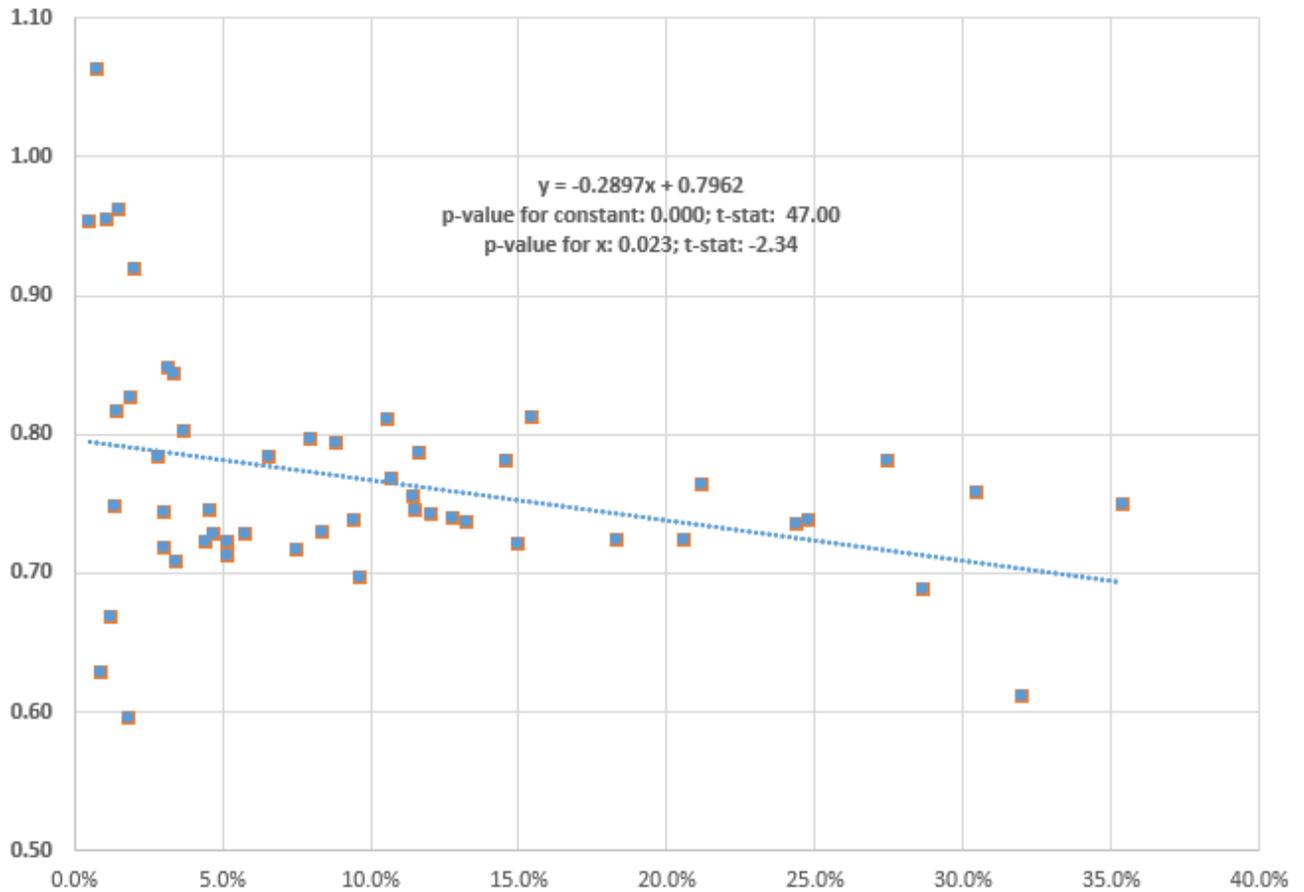

**Figure 1: Scatter Plot and Regression of Earnings Disparity (on the y-axis) and Percentage of Black workers in the workforce (on the x-axis)**

- Date range: January 2017 – December 2019. Source: Earnings Disparities by Race and Ethnicity, based on the Current Population Survey (CPS), July 2020.
- Observation #1: The slope is negative between the earnings disparity and the percentage of Black workers in the workforce, indicating that the percentage of Blacks increases as the earnings disparity grows. Both the coefficient and constant are statistically significant at the 5% level.
- Observation #2: For every 10% increase in the percentage of Blacks in the workforce, the earnings disparity increases by 2.9%.

## 4. CONCLUSIONS

The United States is currently experiencing a period of heightened racial differences pertaining to economic disparities. Absent from the discussion are locations like Idaho that defy the narrative. In Idaho, Blacks earn more than Whites and other races - a remarkable accomplishment. This paper examines the potential factors that might have contributed to this success and explores how Idaho Blacks can serve as a national model community for reducing economic disparities in other parts of the country. While it is challenging to pinpoint the exact reasons for Idaho's Black community's success, preliminary research suggests that factors such as fewer barriers to land ownership, smaller populations, well-knit communities, men's involvement in the family, and a relatively less hostile environment have played a significant role.

Ironically, the history of Idaho Blacks is neither well-documented nor studied in-depth. Historians can uncover valuable insights into what has led to this community's prosperity by further studying Idaho Black's unique situation in the nation's history. A research team effort is required to go through the digital archives of 65 newspapers systematically over the last 150 years. We can then look at how the Idaho factors can be transported and applied to other parts of the country, leading to a more equitable and prosperous future for all Americans.

# 5. COMPLIANCE WITH ETHICAL STANDARDS

**Funding:** This study is not funded by any entity.

**Conflict of Interest:** Authors do not have any conflicts of interest associated with this article.

**Ethical approval:** This article does not contain any studies with human participants or animals.

**REFERENCES (APA format)**